# IMPLEMENTATION AND EVALUATION OF MEASUREMENT-BASED ADMISSION CONTROL SCHEMES WITHIN A CONVERGED NETWORKS QOS MANAGEMENT FRAMEWORK

Suleiman Y. Yerima

School of Computing and Information Engineering, University of Ulster, N. Ireland, United Kingdom.

s.yerima@ulster.ac.uk

## ABSTRACT

*Policy-based network management (PBNM) paradigms provide an effective tool for end-to-end resource management in converged next generation networks by enabling unified, adaptive and scalable solutions that integrate and co-ordinate diverse resource management mechanisms associated with heterogeneous access technologies. In our project, a PBNM framework for end-to-end QoS management in converged networks is being developed. The framework consists of distributed functional entities managed within a policy-based infrastructure to provide QoS and resource management in converged networks. Within any QoS control framework, an effective admission control scheme is essential for maintaining the QoS of flows present in the network. Measurement based admission control (MBAC) and parameter based admission control (PBAC) are two commonly used approaches. This paper presents the implementation and analysis of various measurement-based admission control schemes developed within a Java-based prototype of our policy-based framework. The evaluation is made with real traffic flows on a Linux-based experimental testbed where the current prototype is deployed. Our results show that unlike with classic MBAC or PBAC only schemes, a hybrid approach that combines both methods can simultaneously result in improved admission control and network utilization efficiency.*

## 1. INTRODUCTION

Next generation networks are increasingly being characterized by the convergence of heterogeneous fixed and wireless access network technologies towards interconnectivity with IP-based core networks. Quality of Service (QoS) control and end-to-end management of network resources along the transport plane in such converged domains to enable satisfactory end user Quality of Experience (QoE) poses significant technical challenges, given the complexity introduced by the integration of diverse QoS control mechanisms that may be deployed within each access technology on the end-to-end transport path. Such technical challenges of heterogeneous QoS management in converged networks could be surmounted by deployment of Policy-Based Network Management (PBNM) techniques. PBNM allows for configuration and control of the network as a whole thus eliminating the need to configure and individually manage each network entity [1].

PBNM can simplify administration of complex operational characteristics of a network, including QoS, access control, network security, and IP address allocation [2]. PBNM systems generally use hierarchical structures and will facilitate the management of next generation networks [3]. Moreover, PBNM provides the means for application transparency across existing and emerging access technologies which permit applications to be transport layer-agnostic when deployed [4]. The aforementioned pros of the PBNM approach and its potential to facilitate QoS management in converged next generation networks motivates its adoption as the underlying design paradigm for the QoS management framework presented in this paper.

Also, in this paper we describe the implementation of admission control schemes as an essential building block of our QoS management framework. Measurement-based admission control is implemented within the framework Java-based prototype that has been developed on a Linux-based experimental testbed. This is driven and enabled by the network monitoring and resource control elements within the subsystems of the framework, thus allowing for closed-loop, adaptive and scalable QoS control along the transport plane infrastructure. Furthermore, an experimental analysis of the measurement-based schemes is presented whilst comparing with parameter-based admission control.

One important characteristic common to both measurement-based and parameter-based approaches to is the inherent trade-off between network resource utilization and the conflicting requirement to maintain the QoS of current flows through admission control. It is well known that in most existing measurement or parameter-based admission control schemes, avoiding congestion situations within the network in order to prevent the violation of QoS requirements of admitted flows is usually achieved at the expense of reduced network resources utilization. This implies that the more conservative an admission control scheme is, the lower the overall resource utilization efficiency that is achievable; and for operators this represents a loss of potential revenue. It is therefore highly desirable to develop admission control schemes that can optimize both resource utilization and QoS control within a PBNM system. Thus, in this paper an admission control strategy that achieves improved admission control efficiency (through lower blocking probability) but without sacrificing resource utilization efficiency is also proposed. Unlike traditional methods, the proposed strategy is enabled via a combined measurement and parameter-based approach, which as shown by experiments conducted with real traffic flows on our testbed, results simultaneously in improved admission control and network resource utilization efficiency.

The rest of the paper is organised as follows. Section 2 provides the background and motivation for our contribution while also discussing related work. Section 3 describes the QoS management framework architecture and its constituent entities. Section 4 describes the Java-based prototype implementation including the testbed configuration and the test scenarios. Section 5 discusses the various measurement-based admission control schemes implemented and compared within the PBNM prototype, while experiments and analysis are presented in section 6. Summary, concluding remarks and possible future work are given in section 7.

## 2. BACKGROUND AND MOTIVATION

Admission control methods are important tools for maintaining the QoS of flows within a network by allowing additional flows only if their addition does not disrupt current flows. Hence, admission control (AC) functionality is a critical element of a PBNM-based QoS management infrastructure designed for QoS control in converged networks. Consequently, the underlying schemes within the AC functionality would significantly impact the performance of the QoS management tool, affecting both end-user Quality of experience as well as service provider revenue generation.

Most admission control methods can be classed as either Parameter-Based Admission Control (PBAC) or Measurement-Based Admission Control (MBAC) [5], [6], [7]. PBAC algorithms may be preferred for their ease of implementation within a PBNM tool as they are based on estimates of parameters such as *peak* or *effective* bandwidth usage in the admission request of a new flow rather than actual measurements from the network. Thus, savings in measurement/monitoring overhead may be made by the use of the PBAC approach. However, underutilization of network resources may occur if some flows are frequently idle or transmit at rates below their estimated peak rates, since a new flow may be denied admission despite the network's capability to carry the flow without disrupting the QoS of current flows. PBAC algorithms that employ peak traffic estimates generally suit situations where all admitted flows are simultaneously sending traffic at their peak rates. On the other hand, one limitation of PBAC is the difficulty of accurately characterizing users' traffic in advance especially in diverse environments like converged networks. Moreover, with an increasing variety of new multimedia services emerging as networks converge towards common IP-based core transport, it becomes more tedious and complex to characterize individual traffic flow types a priori. One solution may be to characterize flows within a small well-defined number of classes of aggregate flows. But the drawback of this approach is that such approximations may affect the performance of the PBAC algorithms.

MBAC algorithms use measurements of the current state of the output interface, or the current state of the network to the destination, if available, in order to estimate whether the new flow can be supported [5]. MBACs rely on measurement of actual traffic load and QoS performance in making admission decision [6]. Thus, at the expense of incurred network measurement/monitoring overhead, better network resource utilization (may be achieved compared to the PBAC method) in situations where bursty application flows are present or when not sending at peak rates. MBACs have the ability to offer QoS to users without a priori traffic specification or real-time policing.

Other benefits that MBAC offers to converged networks environment compared to the traditional PBAC include:

1. Adaptability: with MBAC traffic load measurements are taken at certain intervals. Therefore unlike in PBAC, large wastages of resources for an entire session can be avoided as MBAC will be able to adapt more readily to changing network conditions.
2. It eliminates the need for traffic policing by the network provider.
3. The user is relieved from the task of a priori traffic specification.
4. With MBAC, users can be charged for actual data usage allowing for a more flexible charging model.

As mentioned earlier, there is usually a trade-off between network resource utilization and admissibility of an AC algorithm (typically expressed as blocking probability) which also reflects service availability. The more conservative the estimation of current available network resource, the higher the blocking probability for the same admission requests, and hence, the lower the overall network resource utilization. It is therefore desirable that the AC algorithm is designed for optimal performance of both resource utilization and blocking probability. As shown by the implementation and evaluation of the AC schemes within the framework presented in this paper, this is possible through a combination of MBAC with PBAC approach to yield a hybrid scheme that outperforms traditional MBAC-only or PBAC-only methods.

## 2.1 Related work

Despite being a widely studied problem in computer and communications networks since the 90's, admission control has continued to receive due attention with proliferation of heterogeneous networks (differing in scale, complexity and speed) and their recent convergence to accommodate various traffic types with diverse QoS requirements. In [5], the authors compared MBAC algorithms with PBAC based on peak usage and their findings concluded that MBAC provided better network utilization for bursty traffic patterns. Their paper however does not provide results of the blocking probabilities to quantify the utilization vs. availability trade-off in the algorithms. In [16], an online measurement-based admission control scheme is presented. The scheme is applied to VBR video where effective bandwidth is estimated using statistical parameters of aggregate video traffic obtained using linear Kalman filter. The same (Gaussian) effective bandwidth estimation used in [16] is employed in this paper for the MBAC-only scheme in our comparative analysis. Unlike ours, their work concentrated on video traffic using simulation to verify the effectiveness in terms of computational efficiency and accuracy.

In much earlier work, an MBAC scheme is proposed in [17] for dynamic call admission control in ATM networks. The paper uses time window-based measurements during which marginal distributions of cell arrivals for aggregate bandwidth are estimated. Several other MBAC works with measurement processes based on time window such as [18]-[20] have appeared in the published literature since then. Although the focus of our work is different from these previous publications, we also adopt a time window based measurement approach for the MBAC-only schemes used for the comparative analyses undertaken in this paper due to its popularity. Papers [21]-[23] provide more insight into MBAC with [23] focusing on measurement-based estimators. In [6], the discussion was centred on the effectiveness of MBAC and its role in 3G and 4G wireless networks. MBAC methods were found to be beneficial to wireless IP networks considering channel capacity limitations and their highly dynamic nature.

On the other hand, PBNM works related to ours have included descriptions and evaluations of the frameworks mostly using simulation and without addressing admission control, for example: QoS adaptation in DiffServ networks [8]; DiffServ QoS management configuration for military environments [9]; Policy-based routing [10]; VPN management [11]; XML-driven configuration of IMS-based networks [12] to mention a few. However, different from the aforementioned, our contribution focuses on evaluation of MBAC and PBAC admission control schemes within a working Java-based PBNM prototype developed on a Linux-based testbed. The architecture of our PBNM-based QoS management framework for converged networks is described next.

## 3. PBNM-BASED QOS MANAGEMENT FRAMEWORK ARCHITECTURE

This section describes the architecture of the QoS framework designed in our project and implemented using Java within a Linux-based testbed network infrastructure. The QoS framework consists of three logical subsystems including: Resource Management Subsystem (RMS), Measurement and Monitoring Subsystem (MMS) and the Context Management and Adaptation Subsystem (CAS).

Figure 1 shows our framework architecture design with various hierarchical layers as is typical of PBNM systems.  The RMS is the subsystem responsible for co-ordination, control and allocation of resources along the end-to-end transport path of a domain under the framework's administration. The RMS is made up of distributed Access/Core resource brokers that perform the role of policy decision points (PDP) within the PBNM QoS control framework. Access resource brokers oversee edge networks resource allocation while core resource brokers oversee core network resource management. These entities achieve configuration of and control (of traffic) within key elements such as routers, gateways, switches (policy enforcement points, PEP) in the transport plane via lower level entities known as the resource controllers (RC) in our architecture. The MBAC algorithms evaluated in this paper are implemented inside the (core) resource broker element as part of an Admission Control block that utilizes measurements from the MMS while using the policy criteria designed for edge-to-edge admission control within the converged network domain. The policies that are processed by the resource broker PDPs are stored in a central policy repository and provide the high-level control directives which can also be altered at run-time without disrupting the operation of the network. The admission control policies are typically expressed in the ECA (Event, Condition, Action) format using {If … then … else…} rules; but the AC algorithms (which may be MABC or PBAC-based) are the key functionalities that drives the (admission control) policies. Hence the AC algorithms design is critical to the overall performance of the RMS's policy-driven process.

The MMS subsystem of the framework is designed to provide closed-loop adaptive and measurement-based QoS control by feeding the RMS with measurement and monitoring data obtained from key PEPs within the network. The MMS is made up of distributed network monitoring entities NMs, as shown in Figure 1. If centralized monitoring is required, this could be provided by the CMM (central measurement and monitoring entity) which can aggregate measurements from various distributed NMs. The NMs provide the raw measurements for the (Measurement Based) Admission Control functionality. The measurements could be sampled at regular intervals using active and/or passive measurement techniques. The measured samples could be used to calculate parameters such as *mean*, *variance* and *standard deviation* used in some MBAC-based effective bandwidth estimation (see section 5.3). The passive measurement method used in the NM entity utilizes the Simple Network Management Protocol (SNMP) to poll the network entity from which the measurement samples are obtained. The implementation of the NM entity within the Java-based prototype is described in section 4.

The CAS (Context Management and Adaptation Subsystem) is incorporated into the framework design to allow for context-aware resource management through context-driven policies. It consists of distributed CAF (context acquisition function) entities whose task is to interact with end-user in order to extract 'context' information or data that could characterize context such as location, time, speed, user defined contexts etc. The ADS (Adaptation elements) are the lower level counterpart entities of the CAFs that configure and control/adapt traffic within the PEPs. The RMS and MMS are the framework subsystems that are most relevant to the admission control functionality hence we focus our description of the prototype implementation on those aspects in the next section.

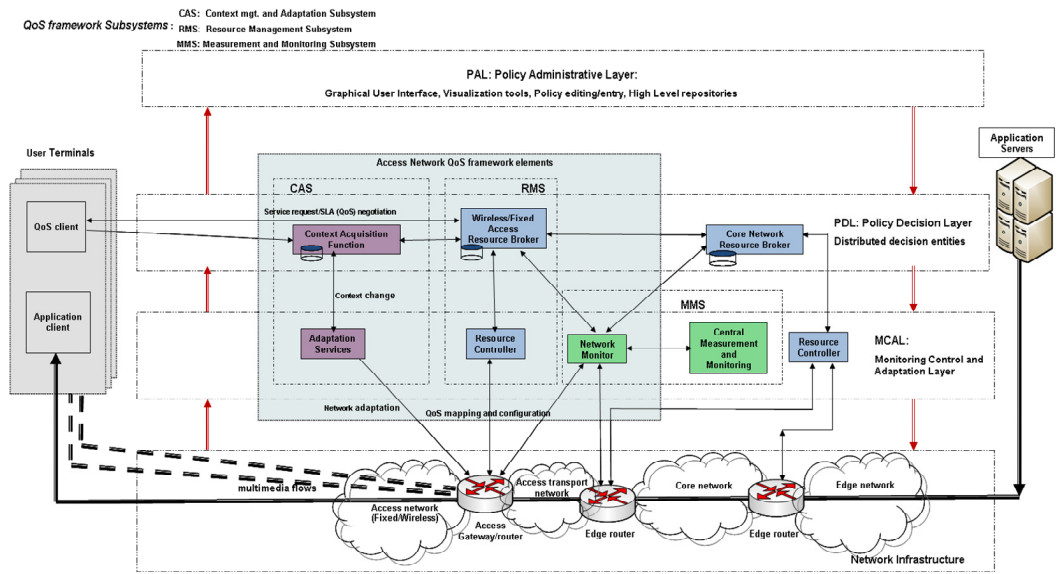

Figure 1: PBNM based QoS framework architecture.

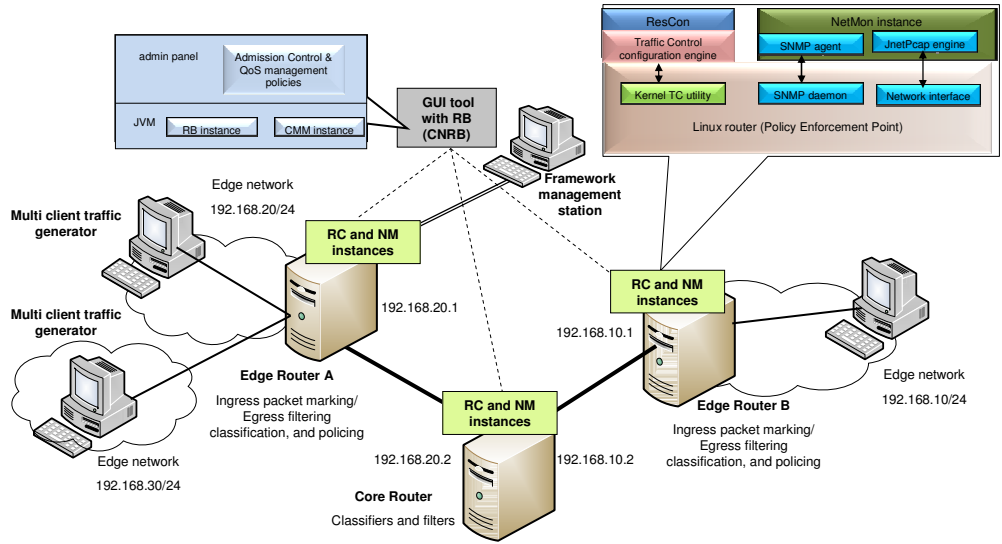

Figure 2: The QoS framework Java-based prototype development on Linux-based testbed.

## 4. FRAMEWORK PROTOTYPE AND TESTBED IMPLEMENTATION

Based on the proposed framework architecture given in the previous section, a working prototype has been built in Java using the open-source *NetBeans* IDE 6.9 platform. Presently, distributed entities of the MMS and RMS subsystems have been implemented on a Linux-based converged network testbed configured as depicted in Figure 2.

### 4.1 Resource Management Subsystem (RMS) implementation

The low level control entity of the RMS is implemented within a ResCon Java class. Distributed ResCon (RC) instances are deployed on the Linux-based routers (Policy Enforcement Points) as shown in Figure 2. The ResCon entities receive configuration instruction via the Java RMI (Remote Method Invocation) communication protocol. These instructions are direct translations of *policy actions* which are triggered within the Resource Broker implementation. A GUI has been implemented for policy editing and issuance of configuration commands via the Resource Broker class. Predefined adaptive polices are also implemented within the RB class such as admission control policies. The MBAC algorithms evaluated in this paper are implemented as separate Java classes that are invoked by the RB class for admission control decisions. Figure 3 depicts a snapshot of the GUI interface to the RB which resides on a separate management station as shown in Figure 2. The *AC* button is used to activate the selected MBAC algorithm class for the QoS administrative domain. As seen from Figure 3, various graphical visualization tools have also been incorporated to provide monitoring capability for various network measurement metrics.

### 4.2 Measurement and Monitoring Subsystem (MMS) implementation

In order to enable closed-loop, adaptive and autonomous QoS control and resource management by the framework prototype, the network monitor element of the MMS class has been implemented as netmon (NM) Java class. Both active and passive measurement capabilities have been implemented within the class. Distributed netmon entities are installed on each PEP (core and edge routers) as shown in Figure 2 and these communicate with the CMM and RB elements in the management station via RMI protocol. The netmon active measurement technique is implemented using packet capture mechanism built using an open-source version of *jNetPcap* [13]. Jnetpcap is a Java library that contains a wrapper for *libcap* library native calls. JNetPcap decodes captured packets in real-time and also provides a large library of network protocols. Furthermore, users can easily add their own protocol definitions using Java SDK. With the packet capture mechanism built in, instances of netmon are able to perform real-time measurements at any network interface within the QoS domain. These measurements are utilized within the MBAC-based classes invoked by the RB for measurement based admission control. The measurements are sampled periodically using the Java timer utility in the netmon implementation. The sampling time can be set via the MBAC-based class of the prototype to suit the specific MBAC algorithm being used. Depending on the MBAC scheme, the effective bandwidth estimation parameters such as mean, standard deviation, variance etc are calculated within a time-window consisting of multiple measurement samples.

In addition to packet capture based measurements, the netmon class also incorporates passive measurement capability via the SNMP (Simple Network Management) protocol. This is achieved by invoking an SNMP agent built using *SNMP4j*, an open source object oriented SNMP API for Java managers and agents [14]. For the experiments conducted in this paper the packet capture based measurements were used because they provided higher sampling rates than the SNMP based method (which is constrained to a 10s interval minimum sampling period).

### 4.3 Experimental testbed configuration

The testbed used for the development of the PBNM framework which implements the AC schemes evaluated in this paper is shown in Figure 2. The framework elements are distributed over the Linux-based machines configured as edge and core routers respectively. Each edge router has a resource controller and a network monitor instance which communicates with the instance of the resource broker (that implements the AC policies) running on the management station. The edge and core routers are built from Intel Xeon 2.66 GHz, 3GB RAM PCs running Ubuntu 10.0.4 Linux. Linux TC (Traffic Control) utility is used to enable packet marking, filtering, classification, policing etc., via the resource controller. netmon measurement samples from the Edge router A are used in the admission control schemes, while traffic is generated from the Ntool's ngen multiclient traffic generator [15].

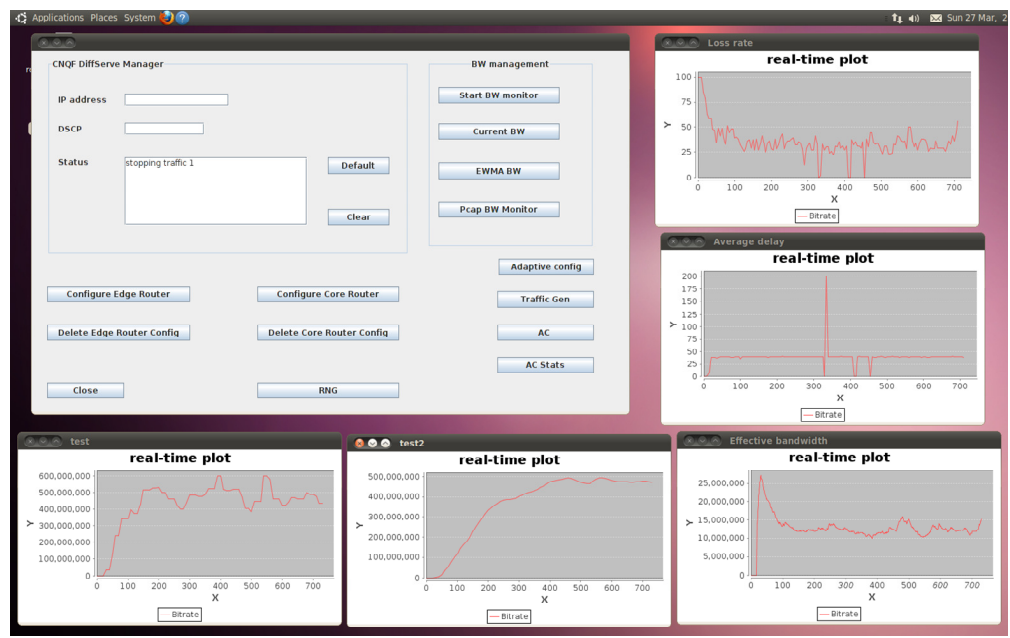

Figure 3: Sample snapshot of an admin and monitoring interface from the current framework prototype on the management station.

## 5. MEASUREMENT-BASED ADMISSION CONTROL

In order to use MBAC methods there needs to be a measurement process [5]. This is enabled by our framework's netmon entity which is driven by a packet capture mechanisms implemented using open-source *jnetpcap* library, as mentioned earlier. MBAC admission decisions in the policy decision entity are made in two steps:

- Obtain (passive/active) measurements of the existing traffic flows in the networks via the *netmon* instance.

- Apply an admission control policy imposed by the QoS management tool to determine whether a new flow should be admitted or rejected.

These two steps correspond to the two distinct logical parts that make up the MBAC mechanism, i.e. the *estimator* and the *admission criteria*:

1. *The estimator:* this is the measurement mechanism which estimates the current network load that is used in the *condition* part of the policies within the PBNM AC block. The measurement parameters used in the estimator are specified within the network monitoring tool or used to configure the *netmon* tool's passive/active measurement process. Hence, the MBAC estimator parameters influence control/management traffic overhead of the PBNM system and should be chosen carefully. The parameters used in the estimator include: *Sampling period, measurement-window size, and average arrival rate/link resource utilization estimation.* From the traffic measurements, we can obtain *mean* and *variance* of traffic flow and estimated link capacity.

2. The *admission criterion is* used within the algorithm that is employed to decide whether to admit or reject a new traffic flow.

Two commonly used MBAC mechanisms also used in schemes implemented within our framework include: (a) *measured sum* and (b) *equivalent/effective bandwidth* [7], [16], [21].

### 5.1 The measured sum approach

The network load of existing traffic is estimated by the measurement entity. The admission criterion for the measured sum is given by:

$$R + C < \theta \cdot T \qquad (1)$$

Where C is the existing network load obtained through measurement while T is the total link capacity and R is the resource (bandwidth) request of the new source. $\theta$ is the utilization target which should be set at a reasonable level because the measurement-based approach will fail if delay variation is exceedingly large, which will occur at high utilization [6]. Measured sum is considered the simplest MBAC algorithm and an adaptation of the simple sum algorithm usually employed in the PBAC scheme for measurement-based environments [6].

### 5.2 Effective bandwidth/equivalent bandwidth approach

An equivalent bandwidth of a set of flows is defined as the bandwidth $C(\varepsilon)$ such that the stationary bandwidth requirement of the set of flows exceeds this value with probability at most $\varepsilon$. Equivalent bandwidth can be calculated using the Gaussian distribution [6], [7], [16]. The Gaussian distribution assumes that instantaneous aggregate arrival rate or current link utilization has a normal distribution. The admission decision is made by checking the sum of the calculated equivalent/ effective capacity (i.e. $C_{eq}$) and the peak rate *p* of a new flow q:

$$C_{eq}(m, \rho, \varepsilon) = m + \alpha \cdot \rho \quad (2)$$

With $\alpha = \sqrt{(2 \ln(1/\varepsilon) + \ln(1/2\pi))}$

Where m is the mean, ρ is the variance while ε denotes the maximum probability with which the sum of new and existing flows is allowed to exceed the link capacity. Figure 4 shows the graph depicting the relationship between ε and α. It implies that higher values of ε will likely result in more conservative estimates of the current bandwidth and hence less aggressive admission control and vice versa; especially with low variation in traffic load (i.e. low variance ρ). Admission criterion is given by equation (1) with $p$ and $C_{eq}$ replacing R and C respectively and θ =1.

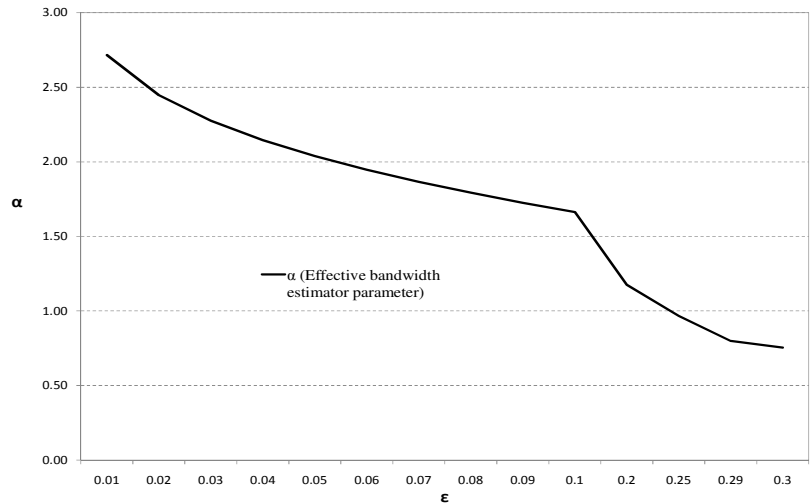

Figure 4: The influence of ε (maximum allowable violation probability) on α (effective bandwidth estimation parameter) in the Gaussian distribution based effective bandwidth estimation approach.

## 5.3 Implemented AC schemes within the QoS Framework Resource Broker

The measurement-based methods that have been implemented within the netmon class/element of the framework prototype are based on the simple sum (for the classic PBAC), the measured sum, equivalent bandwidth (for the classic MBACs), as a well as a hybrid scheme that combines an MBAC and PBAC methods. These implemented algorithms include:

### 5.3.1 The estimated sum PBAC method (PBAC-ES)

Due to its ease of implementation in practical systems, the estimated sum parameter based admission control method provides a commonly used mechanism to estimate the network load in parameter based schemes. The admission criterion is given by:

$$R_2 + C_2 < \theta.T \quad (3)$$

We will call this scheme the PBAC with estimated sum (PBAC-ES). Within the prototype, each arriving flow's peak and average rates are obtained and used to estimate $C_2$ and $R_2$ and equation (3) gives the admission criterion employed. In our experiments, the peak estimates of the flows are used in computing (3) with θ =1. T is the total available link bandwidth.

### 5.3.2 Sliding window measured sum average (SWMSA)

This method is based on the classic measured sum approach. It uses a sliding time window within which the sum of the measured link utilization/available bandwidth is averaged. The measurement window size is set at 10s for the experiments undertaken in this paper while the sampling period is 1s. Both parameters can be changed during run time.

### 5.3.3 Gaussian equivalent bandwidth (GEB)

This method is based on the Gaussian Estimation approach outlined earlier. The admission criterion used within the implemented class used the equivalent bandwidth estimator given in equation (2). With the GEB method, the mean, variance and standard deviation of the measurements taken every X seconds are computed over a time window T such that T/X = 10 (T is currently set to 10 for the experiments in this paper). For the experiments undertaken we used the value $\varepsilon = 0.3$. This corresponds to $\alpha$ being 0.75 approximately (see Figure 4).

### 5.3.4 Weighted average-PBAC approach (EWMA-PBAC)

This uses the weighted sum of the previous estimate and the current measurement to calculate the new estimate of link utilization/available bandwidth. Measurements are taken every second for the analysis in this paper. The method is implemented with an admission criterion:

$$R_3 + C_3 < T \qquad (4)$$

Where $R_3$ is the peak rate of the requesting flow, T is the total bandwidth available, while $C_3$ is the current measurement estimated by:

$$C_3 = \min \{M_t, C_2\} \qquad (5)$$

$$\text{with } M_t = \beta \cdot m + (1-\beta) \cdot M_{t-1} \qquad (6)$$

Where m is the current measured bandwidth usage by all admitted flows and $M_t$ is the current EWMA estimate of the measured bandwidth usage which captures historical data, $\beta$ ($0 < \beta < 1$) is the weight which determines the importance of historical data in $M_t$, and $C_2$ is the estimated link bandwidth usage by simple peak summation of all admitted flows. In the experiments, $\beta = 0.2$ is employed as this is small enough to allow detection in small shifts in current bandwidth usage [5]. Note that this scheme is effectively a hybrid MBAC/PBAC approach allowing it to leverage the individual merits of both methods.

## 6. EXPERIMENTS AND ANALYSIS

A set of experiments were undertaken using the testbed configuration shown in Figure 2. The ingress edge router A of the testbed accepts real traffic flows from various edge networks. Traffic arrival from heterogeneous edge networks were emulated using *ngen* [15] traffic generator which configured the flows with various domain IP addresses representing arrivals from different edge networks. A network monitoring station with the QoS management prototype where the various AC schemes were implemented was connected to the testbed as shown in Figure 2, with instances of the *netmon* network measurement entities installed on the edge and core routers. The MBAC algorithms operate as a component of the QoS framework prototype and within the Resource Broker on the management station and decides whether a flow is accepted or denied once it has been created. The metrics considered for the comparative analyses include the average link utilization on the ingress router interface and the percentage of admitted flows to admission requests (expressed as blocking probability).

## 6.1 Traffic flow configuration

The *ngen* tool is configured to generate traffic with exponentially distributed interarrival intervals. Each flow is terminated after a random exponentially distributed time interval to allow for random parallel flows with different lifetimes during the experiments. A flow is configured to have between minimum average lifetime of 30 seconds and maximum of 120 seconds. Additionally, each flow is generated with a 1Mbps average rate and 1.2 Mbps peak rate. Traffic statistics are gathered and displayed by inbuilt Java-based statistics and graphing tools within the framework.

## 6.2 Edge router configurations

A token bucket filter is configured on the output interface of the ingress routers in order to limit the total output bandwidth to 10 Mbps. The parameters of the TBF configured via Linux Traffic Control utility on the Linux-based edge routers are as follows: Rate: 10 Mbps (10000Kbit), burst 1250000 bytes (1Mbit/sec burst size), buffer size 1250000 bytes, MTU 1540. The Linux TC utility command is shown below:

*sudo tc qdisc add dev eth0 parent 1:0 handle 20: tbf burst 1250000 limit 1250000 mtu 1540 rate 1250000bps*

## 6.3 Result analysis

Figure 5 shows moving average of the network usage over time for the EWMA-PBAC and PBAC-ES schemes in a test run. It is evident from the plots that the EWMA-PBAC hybrid scheme yields better link bandwidth usage than the PBAC only scheme from this figure. This is confirmed by the results in Figure 6. In Figure 5, the moving average of the network usage over time for the EWMA-PBAC and GEB schemes in a test run are depicted. Here also it is evident that the hybrid scheme enables better use of the available link bandwidth from the higher peaks obtained compared to the GEB (which is an MBAC-only scheme).

Figure 5 shows the average utilization (i.e. the average percentage of the total available bandwidth utilized) for arriving requests during the experiments. Ten test runs were made and the average taken to yield the results shown in Figures 7 and 8. The GEB gives lowest average utilization followed by the PBAC-ES and the SWMSA. As expected, the corresponding blocking probabilities from Figure 8 are in the same order from highest to lowest (GEB, PBAC-ES and SWMSA). This is because of the inherent trade-off between resource utilization and blocking probability normally exhibited by traditional single type AC algorithms. Note also that the SWMSA provides better utilization than the PBAC-ES. This can be attributed to the estimation by actual measurement in SWMSA being more accurate than the peak estimates of PBAC-ES (as the flows don't always send packets at their peak rates).

The fact that GEB achieved lower utilization than PBAC-ES and SWMSA confirms the conservative nature in admission control generally associated with Gaussian distribution based AC approaches [7]. On the other hand, GEB is more sensitive to variation in traffic load; higher values of the variance $\rho$ will result in higher estimates of admitted traffic and hence tighter admission control and vice versa. Again, very low values of $\varepsilon$ will tighten admission control even further. Note that value of $\varepsilon = 0.3$ used in the experiments allows for less aggressive admission control and also operates the GEB algorithm at the lowest region of the achievable blocking probabilities and hence is a sensible choice of parameter for comparative evaluation of GEB against the other schemes. Furthermore, the inherent trade-off between utilization and blocking probability implies that lower values of $\varepsilon = 0.3$ will likely result in lower utilization as higher blocking probability of admission requests would be obtained with $\varepsilon$ lower than 0.3.

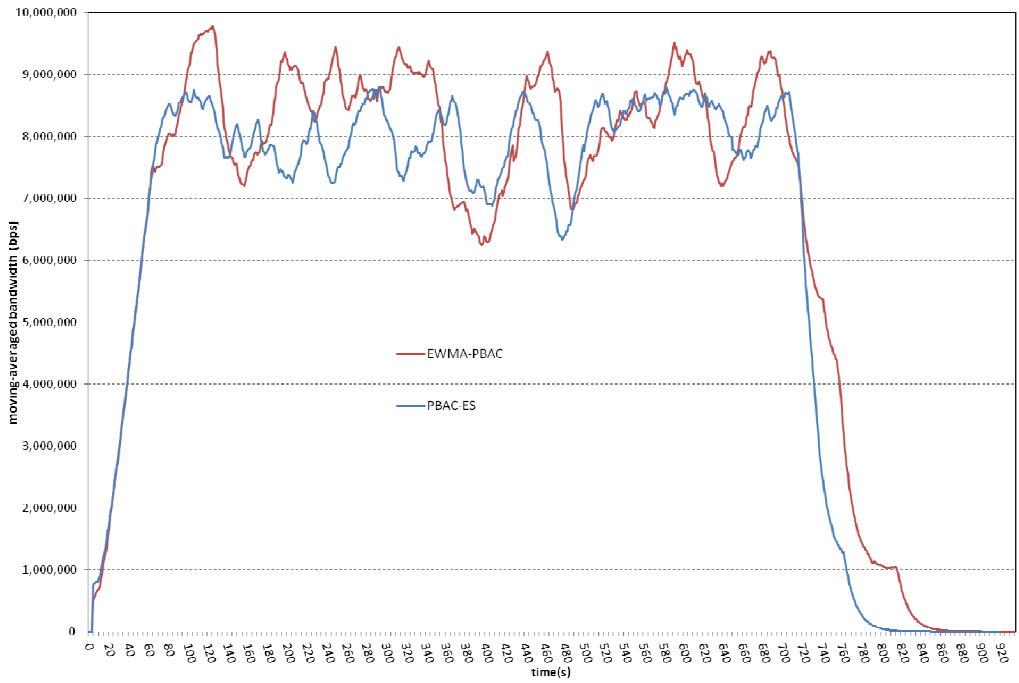

Figure 5: Moving-averaged network usage over time as measured for EWMA-PBAC and PBAC-ES schemes during a test run.

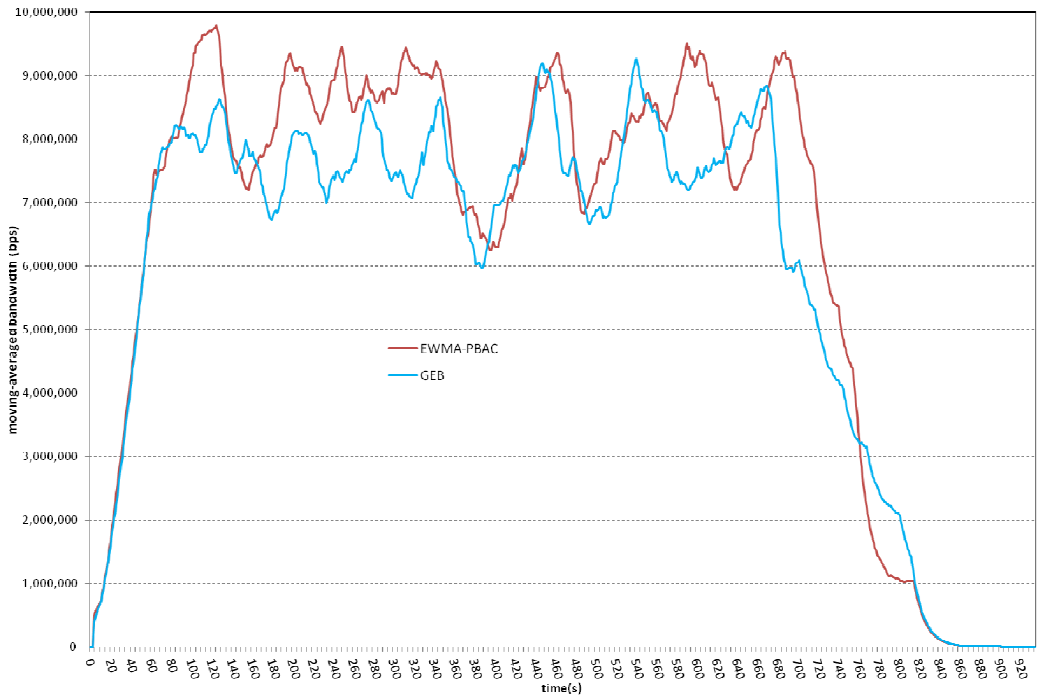

Figure 6: Moving-averaged network usage over time as measured for EWMA-PBAC and GEB schemes during a test run.

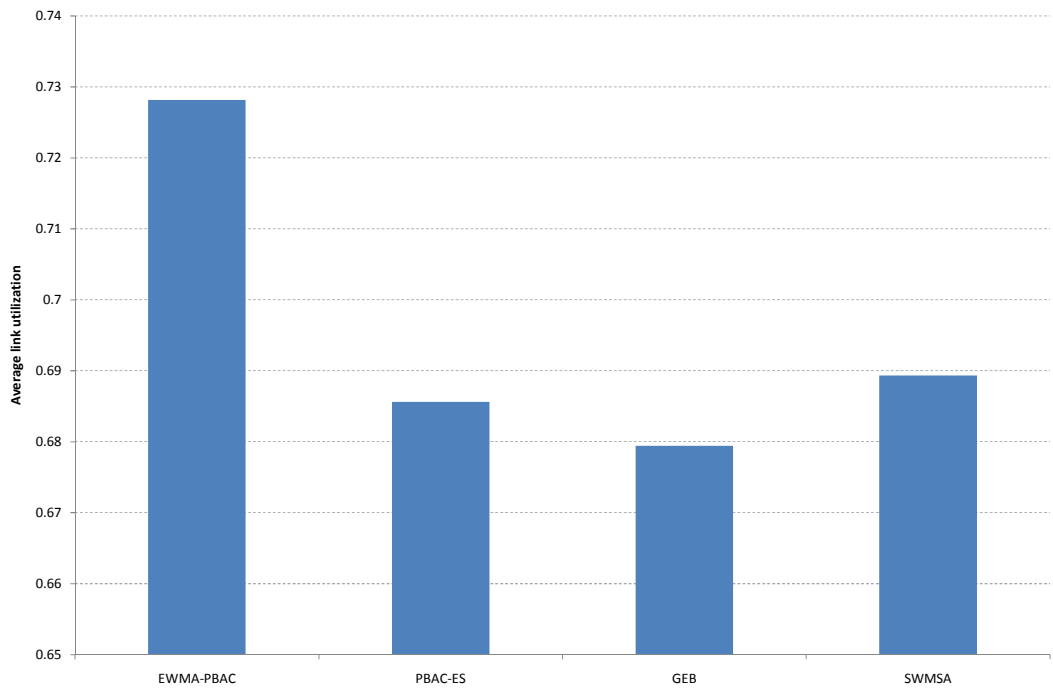

Figure 7: Average utilization

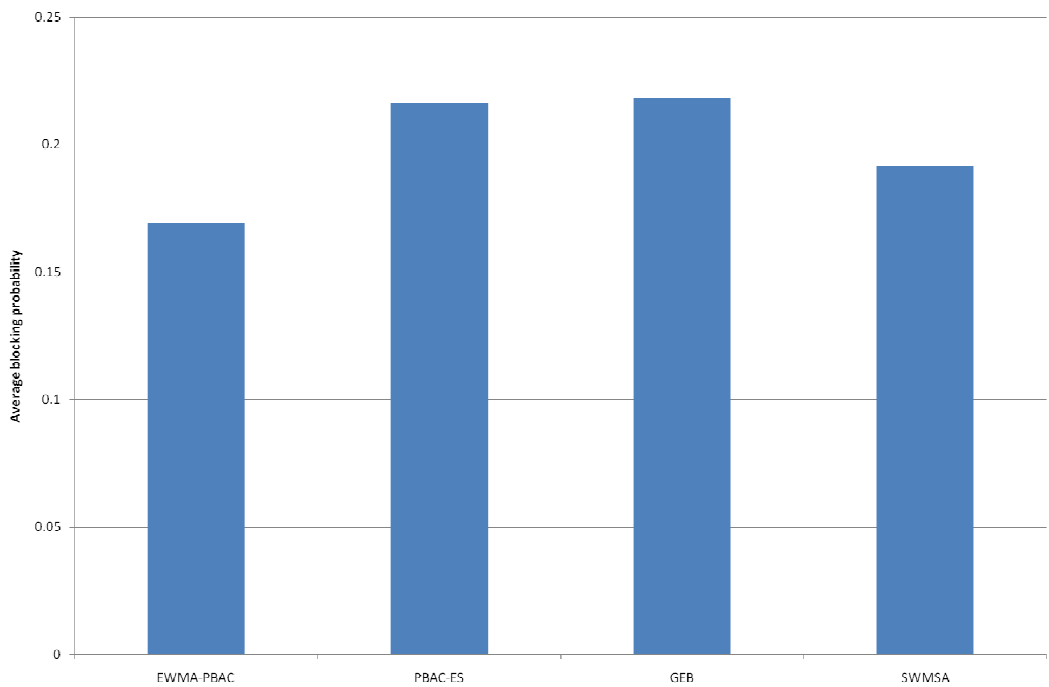

Figure 8: Blocking probability

From Figure 7, it can be observed that the hybrid EWMA-PBAC scheme had the highest average utilization of all the compared schemes but also gave the lowest blocking probability i.e. highest number of admitted flows into the network as shown in Figure 8. The 95% confidence intervals for the results in Figure 8 as derived from Student's t distribution are shown in Table 1. From the table it is clear that the experimental runs undertaken for each scheme on the testbed fell within an acceptable error margin. Hence, we note that the hybrid EWMA-PBAC scheme is the only one that yielded high peak utilization as well as lowest blocking probability thus simultaneously optimizing both.

Table 1. The 95% Confidence Interval values for average blocking probability from 10 test runs for each AC scheme computed with Student's t distribution.

|  | EWMA-PBAC | PBAC-ES | GEB | SAWMSA |
|---|---|---|---|---|
| Average blocking probability | 0.1691 | 0.2164 | 0.2182 | 0.1918 |
| Confidence Interval | ±0.0140 | ±0.0146 | ±0.0178 | ±0.0154 |

Table 2 summarizes the gain in performance that accrues from the EWMA-PBAC approach relative to the other single-method traditional schemes. The use of EWMA-PBAC within the PBNM framework yielded a 22.5% decrease in blocking probability over GEB and 11.8 % decrease over SWMSA. To put this into perspective, this implies that 22.5% more requests will be granted resulting in about 7.2% increase in average utilization by the choice of EWMA-PBAC over GEB. Table 2 shows that 21.9% more requests were granted with consequent 6.2% increase in utilization when EWMA-PBAC is used instead of PBAC-ES; while 11.8% more flows are likely to be admitted with 5.6% increase in utilization over SWMSA. Thus, from the results obtained, we can conclude that the hybrid scheme is able to overcome the situation where the commonly encountered trade-off between utilization and blocking probability exists with single classical PBAC only or MBAC only schemes. Consequently, more admission requests from arriving flows would be granted leading to potential revenue increase.

Table 2. Performance gain achieved by the EWMA-PBAC over the other traditional single-method schemes.

|  | PBAC-ES | GEB | SAWMSA |
|---|---|---|---|
| Increase in average utilization achieved by EWMA- PBAC | 6.2% | 7.2% | 5.6% |
| Decrease in blocking probability achieved by EWMA- PBAC | 21.9% | 22.5% | 11.8% |

## 7. CONCLUSION AND FUTURE WORK

MBAC and PBAC are two commonly used approached for admission control. MBAC methods are known to perform better when bursty flows are present as they rely on actual measurements from the network. On the other hand PBAC methods are easier to implement and could perform well when traffic is sent at peak rates but have a requirement for a priori specification of traffic by the user. We implemented and compared representative PBAC and MBAC only schemes to a hybrid scheme proposed in this paper. The investigation was undertaken with real-traffic on a Linux-based testbed where the AC schemes have been implemented within a policy-based QoS management prototype built with Java. As shown by the results of the experiments, MBAC and PBAC schemes inherently achieve better network resource utilization at the expense of higher blocking probability which translates to reduced service availability. The results obtained also showed the proposed EWMA-PBAC approach was capable of outperforming the others in terms of network resource utilization but without sacrificing service availability as illustrated by the lower blocking probability performance compared to the others. This indicates the potential for hybrid PBAC-MBAC approach to maximize both resource utilization and service availability simultaneously. In future work, it is intended to extend the algorithms for class-based AC analysis using the testbed and prototype. Furthermore, implementation of Machine Learning based decision engines to leverage the MBAC schemes implemented for further optimization of the efficiency of the AC functionality within the PBNM prototype will be investigated.

**Authors**

**Suleiman Y. Yerima** is currently a Postdoctoral Researcher at the School of Computing and Information Engineering, University of Ulster, N. Ireland, United Kingdom. He was previously a researcher at the Integrated Communications Research Centre (ICRC), University of Glamorgan, UK where he obtained a PhD in Wireless Communications and Networking in May 2009. He holds an MSc (With Distinction) in Personal, Mobile and Satellite Communications from University of Bradford, UK (2004) and a B.Eng. degree in Electrical and Computer Engineering from Federal University of Technology, Minna, Nigeria (2000). He was the best student paper award recipient in the World Congress of Engineering (WCE) International Conference of Wireless Networks held in London, June 2009. His current work is focused on machine learning based network management in converged next generation networks and application of machine learning in teletraffic analysis. He is also a member of IEEE, IET and IAENG societies.